\documentclass[aps,pra,twocolumn, 10pt, showpacs,superscriptaddress,groupedaddress]{revtex4-1}  

\newcommand{\ket}[1]{\left\vert#1\right\rangle}

\newcommand{\Rb}{$^{87}\text{Rb }$}

\newcommand{\eq}[2][{}]{\begin{equation}\label{#1} #2 \end{equation}}

\newcommand{\eg}{\emph{e.g.} }

\usepackage{graphicx}
\usepackage{epstopdf}
\usepackage{bbold}
\usepackage{comment}
\usepackage{color}
\usepackage{soul}
\usepackage[abs]{overpic}
\usepackage{amsmath}
\usepackage{lipsum}
\usepackage{hyperref}
\usepackage{setspace}
\begin{document}

\author{Jonathan Coslovsky}
\affiliation{Department of Physics of Complex Systems, Weizmann Institute of Science, Rehovot 76100, Israel }
\author{Gadi Afek}
\affiliation{Department of Physics of Complex Systems, Weizmann Institute of Science, Rehovot 76100, Israel }
\author{Alexander Mil}
\altaffiliation[Current address: ]{Kirchhoff-Institut f\"ur Physik, Ruprecht-Karls-Universit\"at Heidelberg, Im Neuenheimer Feld 227, 69120 Heidelberg, Germany }
\author{Ido Almog}
\affiliation{Department of Physics of Complex Systems, Weizmann Institute of Science, Rehovot 76100, Israel }
\author{Nir Davidson}
\affiliation{Department of Physics of Complex Systems, Weizmann Institute of Science, Rehovot 76100, Israel }

\title[]{
Spectroscopic Measurement of the Softness of Ultra-Cold Atomic Collisions}

\begin{abstract}

The softness of elastic atomic collisions, defined as the average number of collisions each atom undergoes until its energy decorrelates significantly, can have a considerable effect on the decay dynamics of atomic coherence. In this paper we combine two spectroscopic methods to measure these dynamics and obtain the collisional softness of ultra-cold atoms in an optical trap: Ramsey spectroscopy to measure the energy decorrelation rate and echo spectroscopy to measure the collision rate. We obtain a value of 2.5 (3) for the collisional softness, in good agreement with previously reported numerical molecular dynamics simulations. This fundamental quantity was used to determine the $s$-wave scattering lengths of different atoms but has not been directly measured. We further show that the decay dynamics of the revival amplitudes in the echo experiment has a transition in its functional decay. The transition time is related to the softness of the collisions and provides yet another way to approximate it. These conclusions are supported by Monte Carlo simulations of the full echo dynamics. The methods presented here can allow measurements of a generalized softness parameter for other two-level quantum systems with discrete spectral fluctuations.

\end{abstract}

\maketitle

Elastic collisions are of great importance in atomic physics, both from a theoretical and a practical point of view. They are relevant for atomic clocks, metrology, quantum information, evaporative cooling, atom-ion hybrid systems and more~\cite{Gibble1991, Gibble1993, Anderson1995, Swallows2011, Meir2016, Hohmann2016Arxiv}. Collisions may also have a significant effect on the coherence properties of an ensemble of atoms, providing either elongation~\cite{Dicke1953, Schmidt1973, berman1975, Rothberg081984, firstenberg2013colloquium, Firstenberg2007, Shuker2007, Sagi2010_Spectrum, Sagi2010_Universal, Sagi2011PRA} or shortening~\cite{carr1954effects,Deutsch2010} of the atomic coherence time.

Considering a rapid collisional process compared to other dynamical timescales~\footnote{the collision time can be estimated by the range of atomic interaction divided by the relative velocity of the colliding atoms. For dilute ultra-cold atoms and excluding mean-field interactions the collision time is on the nanosecond scale, much shorter than the mean time between collisions and the oscillation time in the trap ($\gtrsim10$~msec and $\gtrsim1$~msec, respectively for our experiment).}, there exist two extremities for a colliding atom in the center-of-mass frame of the interacting ensemble: ``hard collisions'', in which the energy of the atom is completely randomized after a single collision, and ``soft collisions'' in which the atomic energy remains almost unchanged after each collision~\cite{Sagi2010_Spectrum}. We therefore define the \textit{``collisional softness''} parameter, $s$, as the number of times an atom has to collide in order for the correlation between its initial and final energies to drop to $1/e$~\footnote{This definition of $s$ is related to the strength parameter of velocity changing collisions, $\alpha$, defined in~\cite{berman1975}, by $s = \frac{1}{1-\alpha^2}$}. The collisional softness of hard collisions is one, since the energy correlation drops to zero after a single collision. Collisions are considered ``soft'' if their softness parameter is much larger than unity. Even though the $s$-wave collisional process considered here is itself is of universal nature, the softness of the collisions can be affected by the confining potential. This can be intuitively understood by considering that only the kinetic energy changes due to a collision whereas the potential energy does not, carrying a ``memory'' of the energy prior to the collision.

More formally, an ensemble of colliding trapped thermal atoms has two relevant characteristic rates. First, the atomic collisions, treated as a Poisson process energy-randomization events, occur at an \textit{average collision rate} $\Gamma_\text{coll}$. Second, the single atom temporal energy autocorrelation function, averaged over the atomic ensemble, decays exponentially with an \textit{energy decorrelation rate} $\Gamma$. The collisional softness is then defined as:
\eq[eq:softness_definition]{s = \Gamma_\text{coll}/\Gamma.}
Collisions with $s=1$ are \textit{hard}, and collisions with $s \gg 1$ are \textit{soft}.

The softness of $s$-wave elastic collisions of ultra-cold bosonic atoms trapped in a harmonic potential and far from a Feshbach resonance was evaluated using molecular dynamics simulations and found to be 2.5~\cite{Monroe1993, Newbury1995, Foot1996, Dalibard1997, Jin1999, dalibard1999collisional, Foot2000,Jin2014}. This value of the softness has been used to determine the elastic collision cross sections of different atoms~\cite{Monroe1993, Newbury1995, Dalibard1997,Foot2000}, but has not been measured directly. For a given number of collisions the softness is the number of collisions required for thermalization in a perturbed trap, having immediate repercussions on the physics of evaporative cooling~\cite{Monroe1993,Foot1996}.

In this paper we present a direct spectroscopic measurement of the softness of ultra-cold atomic collisions. We do so using a combination of two spectroscopic methods (Ramsey and echo~\cite{Hahn}), in two opposite regimes of low and high collision rates. First we show that the coherence of an atomic ensemble in an echo experiment at low density asymptotically depends only on $\Gamma_\text{coll}$. We then show that in a high density Ramsey experiment the decay is independent of the collision rate and can be fitted to reveal the energy decorrelation rate $\Gamma$. By combining these two measurements, we are able to quantitatively extract the $s$-wave collisional softness of cold $^{87}$Rb atoms in an optical dipole trap. We obtain good agreement with previously reported theoretical results from molecular dynamics simulations, validating our method and laying the foundations for its application in measuring the softness of other collisional processes.

We further show that the coherence decay in an echo experiment is qualitatively different for short and long times~\cite{Schmidt1973,berman1975}. This can be used to approximate the softness by combining a Ramsey measurement and an echo measurement in a single, intermediate density regime. These methods may allow measurements of a softness parameter for other two-level quantum systems that have discrete energy fluctuations~\cite{Dicke1953,koch2007model, wodkiewicz1984noise,ambrose1991fluorescence,uren19851}.

\subsection*{Spectroscopic signatures of collisional softness}

\Rb atoms trapped in an optical dipole trap experience a differential AC-Stark shift imposed by the different detuning of the trapping laser from their two ground state hyperfine levels. If the mean time between atomic collisions is larger than the oscillation period in the trap, the fast oscillations can be averaged. The rate of the phase accumulated by the wavefunction, determined by the detuning, then depends on the average energy~\cite{Kuhr2005}. Effectively this creates a stationary inhomogeneous broadening of the spectrum, decreasing the coherence time of the ensemble. 

Due to this effect, the dynamics of the hyperfine coherence in a Ramsey ($\pi/2-\pi/2$) experiment with no collisions is given by
\eq[eq:Meschede]{C_\text{R}(t) = \left[1+0.95(t/\tau)^2\right]^{-3/2}}
for an ensemble of two-level atoms in thermal equilibrium in an optical harmonic potential~\cite{Kuhr2005}. The bare Ramsey time is given by $\tau\approx2\hbar/\eta k_BT$. Here $T$ is the temperature of the cloud and $\eta$ is the ratio between the hyperfine splitting and the detuning of the trapping laser~\footnote{For $^{87}$Rb and a YAG 1064~nm trapping laser $\eta\approx7\times10^{-5}$}. In an echo experiment ($\pi/2-\pi-\pi/2$), where the echo-pulse is given at time $t_\pi$, the echo coherence $C_\text{E}(2t_\pi)\equiv C(t=2t_\pi)$ fully revives in the absence of elastic atomic collisions due to the stationarity of the trap perturbation~\footnote{For strong trap perturbations and for chaotic traps this may not be the case~\cite{Andersen2003,Andersen2006}}.

Factoring in the effect of elastic atomic collisions, the Ramsey and echo coherences have a complicated behaviour~\cite{firstenberg2013colloquium}. However, they both have some useful, simple limits. For high density $n_\text{high}$, the spectrum is collisionally-narrowed, resulting in an elongated Ramsey coherence, that can be approximated by the generalized Gumbel function~\cite{Sagi2010_Universal}
\eq{C_\text{R}(t) \sim \exp \left[-\frac{2.86}{\Gamma^2\tau^2}\left(e^{-\Gamma t}+\Gamma t-1\right)\right],\label{eq:Gumbel}}
dependent only on $\tau$ and the energy decorrelation rate $\Gamma$, and not on the collision rate $\Gamma_\text{coll}$.

In the opposite regime, of low density $n_\text{low}$, the asymptotic long-time echo coherence behaves as 
\eq[eq:low_density_corr]{C_\text{E}(2t_\pi) \sim \exp(-2\Gamma_\text{coll}t_\pi).} 
In this regime the coherence depends solely on the collision rate $\Gamma_{\text{coll}}$ and not on the energy decorrelation rate $\Gamma$. This is due to the fact that every collision, no matter how soft, causes a finite deflection in the atomic trajectory. As $t\to\infty$ this deflection will fully decohere the atom. This implies that the coherence in this long-time regime is nothing but the fraction of atoms that did not collide.

\begin{figure}
\centering
\begin{overpic}
[width=\linewidth]{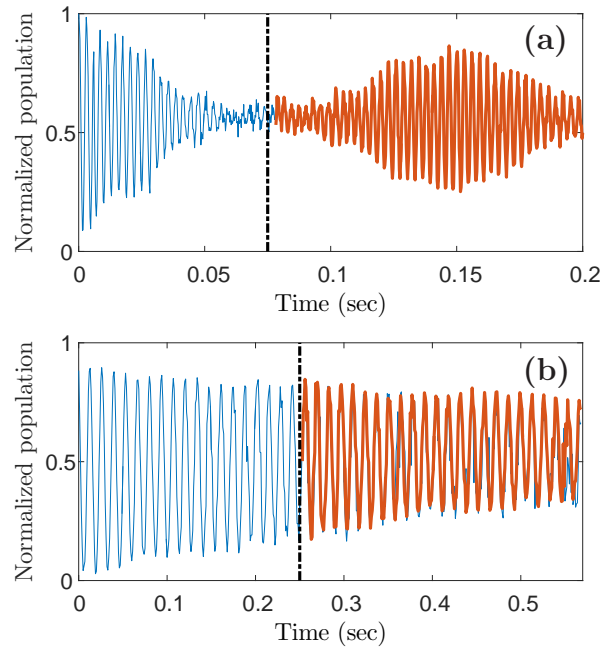}
\put(200,230){\large \textbf{(a)}}
\put(200,105){\large \textbf{(b)}}
\end{overpic}
\caption[Echo rephasing.]{Comparison between echo revival in the regimes of (a) low collision rate ($\Gamma_\text{coll} \tau \approx 0.15$) and (b) high collision rate ($\Gamma_\text{coll} \tau \approx 10$)~\cite{Note5}. The Ramsey signal without the application of the echo (thin, blue) is compared to the echo signal after the application of the pulse (thick, red). Black dashed lines represent the times of the application of the $\pi$-pulses. The envelope of the obtained Ramsey fringes is an indication of the atomic coherence. (a) At low collision rates the amplitude of the echo revival is high. The value of the coherence at the peak of the revival, $C(t=2t_\pi)$, is defined as the echo coherence $C_\text{E}(2t_\pi)$. (b) At high collision rates the echo pulse essentially has no effect on the coherence. The vertical axis represents the fraction of atoms in the excited state after the last Raman pulse.}
\label{fig:fig1}
\end{figure}

Fig.~\ref{fig:fig1} illustrates the role of collisions in an echo experiment. It presents the normalized upper hyperfine state population as a function of the time between the Ramsey pulses. In (a) the measurement is performed with very low collision rate ($\Gamma_\text{coll} \tau \approx 0.15$)~\footnote{Low/high collision rates are defined, throughout this paper, with respect to the bare Ramsey time $\tau$. In both regimes the collision rate is much smaller than the trapping frequencies.}. The coherence decays fast, and the echo revival amplitude is significant. On the other hand, in (b) the collision rate is much higher ($\Gamma_\text{coll} \tau \approx 10$). The Ramsey decay is slower (partially due to collisional narrowing) and the echo revival is negligible, manifesting the failure of the echo due to atomic collisions.

Both the collision rate and the energy decorrelation rate are proportional to the multiplication of the atomic density $n$ with the average velocity $v_\text{rel} \sim \sqrt{T}$: $\Gamma_\text{coll} = n \sigma v_\text{rel}$. Therefore, knowing the ratios $n_\text{high}/n_\text{low}$ and $T_\text{high}/T_\text{low}$ at the two different experimental conditions allows for the normalization of the collision rate, measured at low density, and the energy decorrelation rate, measured at high density, and extraction of the collisional softness
\eq{s = \left(\frac{n_{\text{high}}}{n_{\text{low}}}\sqrt{\frac{T_{\text{high}}}{T_{\text{low}}}}\right)\frac{\Gamma_{\text{coll}}^\text{low}}{\Gamma^\text{high}}. \label{eq:getting_softness}}

We note that in the intermediate regime $\Gamma \tau \approx 1$ it was theoretically shown that the spectrum of an ensemble of colliding atoms depends weakly on the softness~\cite{Sagi2010_Spectrum}. It was further suggested that it may be possible to distinguish between the two extreme cases of hard and soft collisions, by measuring a Dicke narrowed spectrum in a Ramsey experiment. Practically, this task turns out to be challenging. Small uncertainties in the experimental conditions, such as the collision rate and the inhomogeneous broadening of the spectrum, may cause an incorrect model to fit well to the experimental data~\cite{Rothberg081984,firstenberg2013colloquium}.

\subsection*{Measuring the collisional softness}

Our apparatus is described in detail in~\cite{Sagi2010_Spectrum}. Briefly, the experiment consists of $^{87}$Rb atoms trapped in a 1064~nm far-detuned crossed-beam optical dipole trap. The atoms are evaporatively-cooled down to two distinct regimes: high density with $n_\text{high} = 3.5~(2)\times10^{12}$~cm$^{-3}$ and $T_\text{high} = 0.56~(2)~\mu$K, and low density with $n_\text{low} = 3.6~(3)\times10^{11}$~cm$^{-3}$ and $T_\text{low} = 6.8~(3)~\mu$K~\footnote{The temperature is measured using time-of-flight and the peak atomic density using $n=\omega_x\omega_y\omega_zN\left(\frac{m}{2\pi k_BT}\right)^{3/2}$, where $\omega_i$ are the trap frequencies, $N$ is the total number of atoms, $m$ is the atomic mass, $k_B$ is the Boltzmann constant and $T$ the measured temperature}. Measurement of the total number of atoms, and hence the peak density, is susceptible to common systematic errors and obtaining an exact value for it is challenging~\cite{Weller2012}. However, as our method relies only on the knowledge of the ratio between densities [Eq.~\eqref{eq:getting_softness}], systematic errors are common-mode rejected. All errors stated in throughout the paper represent a 1$\sigma$ confidence level.

The coherence is measured between the first-order Zeeman insensitive hyperfine $\ket{1} \equiv \ket{F=1,m_F=0}$ and  $\ket{2} \equiv \ket{F=2,m_F=0}$ states of the $5^2S_{1/2}$ manifold. The atoms are prepared by optical pumping and microwave transitions in state $\ket{1}$. We then use a microwave $\sim 6.8$~GHz control to perform Ramsey ($\pi/2-\pi/2$) or phase-scanned echo ($\pi/2-\pi-\pi/2$) manipulations on the atoms. At the end of each experiment we use a state-selective fluorescence-detection scheme to evaluate $N_2/(N_1+N_2)$, the fraction of atoms at state $\ket{2}$. The coherence is defined as the normalized amplitude of the fringes of the Ramsey and echo data.

We measure the collisional softness using Eq.~\eqref{eq:getting_softness}, by first obtaining $\Gamma_\text{coll}^\text{low}$ from an echo measurement in the low density regime by fitting the asymptotic decay described by Eq.~\eqref{eq:low_density_corr} and then obtaining $\Gamma^\text{high}$, given by Eq.~\eqref{eq:Gumbel} from a Ramsey measurement in the opposite regime.

\begin{figure}
\centering
\begin{overpic}
[width=\linewidth]{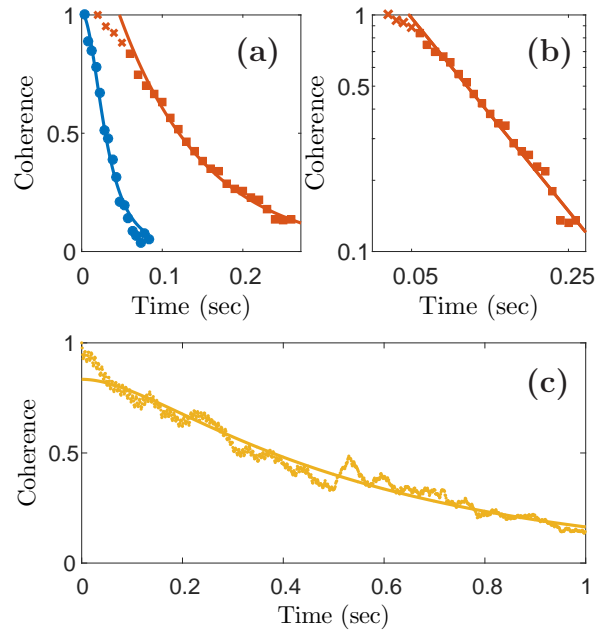}
\put(90,225){\large \textbf{(a)}}
\put(200,225){\large \textbf{(b)}}
\put(200,100){\large \textbf{(c)}}
\end{overpic}
\caption[{Simulations of collisional narrowing with hard and soft collisions.}]{{Ramsey and echo experiments.} (a) The atomic coherence as a function of time in Ramsey (blue circles) and echo (red squares) experiments at low density. The decay of the Ramsey signal yields a coherence time of $\tau_\text{low} = 37$~(1)~ms. (b) A linear fit to the tail of the echo decay, in logarithmic scale, gives $\Gamma_{\text{coll}}^\text{low} = 9.4~(3)~\text{s}^{-1}$. Crosses are short-time data points excluded from the fit. Solid lines represent the fitted functions (see text). Time in the echo experiment corresponds to $2t_\pi$. (c) Ramsey measurement of the energy decorrelation rate $\Gamma^\text{high}$ at high density. The measured coherence is fit to a Gumbel function [Eq.~\eqref{eq:Gumbel}] with the energy decorrelation rate as a fitting parameter. This yields $\Gamma^\text{high} = 10.6~(1)~\text{s}^{-1}$.}
\label{fig:fig2}
\end{figure}

Focusing first on the low density regime, the resulting echo coherence and an additional Ramsey measurement at the same experimental conditions are presented in Fig.~\ref{fig:fig2}(a). The echo decay time is indeed much longer than that of the Ramsey experiment (by about a factor of four). The extracted low-density bare Ramsey time is $\tau_\text{low} = 37~(1)$~ms, compared to 32~(1)~ms obtained directly from the measured temperature. The echo measurement exhibits a long-time linear decay in a semi-logarithmic scale [Fig.~\ref{fig:fig2}(b)], confirming the expected exponential decay of Eq.~\eqref{eq:low_density_corr}. The slope, excluding short times, gives a collision rate of $\Gamma_{\text{coll}}^\text{low} = 9.4~(3)~\text{s}^{-1}$.  
Next we obtain the energy decorrelation rate, $\Gamma^\text{high}$, from the collisional narrowing of a high-density Ramsey measurement. The atomic coherence is shown in Fig.~\ref{fig:fig2}(c). From the measured temperature, we expect $\tau_\text{high} = 390\ (15)$~ms. We use this value as a fixed parameter and fit the coherence data to Eq.~\eqref{eq:Gumbel}, extracting $\Gamma^\text{high} = 10.6~(1)~\text{s}^{-1}$.

The softness is then calculated using Eq.~\eqref{eq:getting_softness} to be $s = 2.5~(3)$~\footnote{This value is correct for harmonic trapping potential that describes well our crossed Gaussian beam optical potential. For other trap shapes and energy distributions it may vary, \eg for a flat box potential $s=1.5$~\cite{Foot1996}}, in excellent agreement with molecular dynamics simulations~\cite{Foot1996}.

\begin{figure}
\centering
\begin{overpic}
[width=\linewidth]{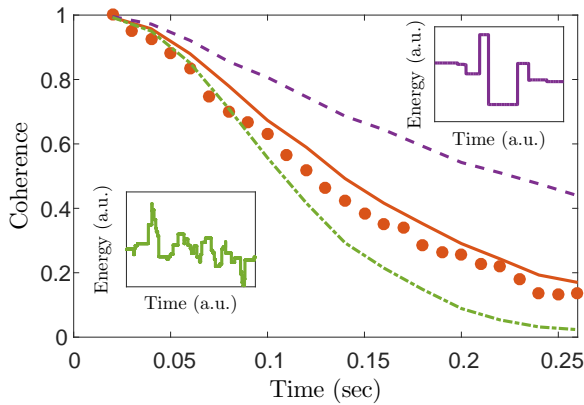}
\end{overpic}
\caption[{Simulations of collisional narrowing with hard and soft collisions.}]{Echo experimental results (circles) compared to Monte Carlo simulations for atoms with an energy distribution corresponding to a 3D harmonic potential with hard ($s=1$, dashed, purple), soft ($s=10$, dash-dot, green) and $s=2.5$ (solid line, red) collisions. The simulation uses the calculated bare Ramsey time of $\tau_\text{low} = 32$ ms, and a constant energy decorrelation rate of the measured $\Gamma^\text{high} = 10.6~\text{s}^{-1}$, rescaled using Eq.~\eqref{eq:getting_softness}. Insets illustrate the energy of a sample atom as a function of time for hard ($s=1$, purple, top right) and soft ($s=10$, green, bottom left) collisions, with the same energy decorrelation rate, $\Gamma$.}
\label{fig:fig3}
\end{figure}

We perform Monte Carlo simulations to study the effect of the softness of the collisions on the full dynamics of the echo decay. The simulation calculates the ensemble coherence of $2\times10^4$ two-level atoms with the energy distribution corresponding to a 3D harmonic potential~\cite{Kuhr2005} as a function of time. The collision rate $\Gamma_\text{coll}$ is drawn from a Poisson distribution and the collisional softness $s$ is generated by introducing controlled correlations between the energy jumps of successive collisional events using the Cholesky decomposition method of the required correlation matrix~\cite{Hardoon2004}. Typical energy trajectories for $s=1$ and $s=10$ are illustrated in the insets of Fig.~\ref{fig:fig3}. A comparison between the experimental data and simulation results is presented in Fig.~\ref{fig:fig3} for the echo experiment with hard ($s=1$), soft ($s=10$) and $s=2.5$ collisions, using the measured energy decorrelation rate, rescaled using Eq.~\eqref{eq:getting_softness}, $\Gamma^\text{low} = \left(\frac{n_\text{high}}{n_\text{low}}\sqrt{\frac{T_\text{high}}{T_\text{low}}}\right)^{-1}\Gamma^\text{high} = 3.76~\text{s}^{-1}$. The simulation of the $s=1$ and $s=10$ clearly disagrees with the data, whereas the $s=2.5$ collisions agrees best with the experimental data for all times and with no fit parameters. The echo decays of the hard and of the soft collisions, with the same energy decorrelation rate, clearly do not agree with the experimental results. 

\subsection*{Measuring the softness using a transition in the functional decay of the echo dynamics}

In the low density regime of $\Gamma\tau\ll 1$, valuable information can be extracted by observing the entire dynamics of the decay of the echo coherence. Eq.~\eqref{eq:low_density_corr} gives the long-time limit of the coherence $C_\text{E}(2t_\pi) \sim \exp\left(-2\Gamma_\text{coll}t_\pi\right)$, depending solely on the collision rate $\Gamma_\text{coll}$. The short-time limit, however, depends on the energy decorrelation rate $\Gamma$ and the bare Ramsey time $\tau$ and is given by~\cite{carr1954effects}-
\eq[eq:low_density_short_time]{C_\text{E}(2t_\pi) \sim \exp\left[-\frac{\Gamma \left(2t_\pi\right)^3}{6\tau^2}\right].}
In both limits the echo coherence decays as $C_\text{E}(2t_\pi)\sim \exp\left(-\beta(2t_\pi)^\alpha\right)$, with $(\alpha,\beta)=\left(1,\Gamma_\text{coll}\right)$ for long times or $(\alpha,\beta)=\left(3,\Gamma/6\tau^2\right)$ for short times. Defining $t_{\text{tr}}\equiv \tau/\sqrt{s}$, a transition time between the two regimes, an interpolating function can be written to describe the entire dynamics, which is the accurate solution in the limit of soft collisions~\cite{berman1975}:
\eq{C_\text{E}(2t_\pi) \sim \exp \left\{-2\Gamma_{\text{coll}}t_\pi\left[1-\sqrt{\frac{\pi}{2}}\frac{t_\text{tr}}{2t_\pi}\text{erf}\left(\frac{\sqrt{2}t_\pi}{t_\text{tr}}\right)\right]\right\}.\label{eq:Transition_func}}

\begin{figure}
\centering
\begin{overpic}[width=\linewidth]{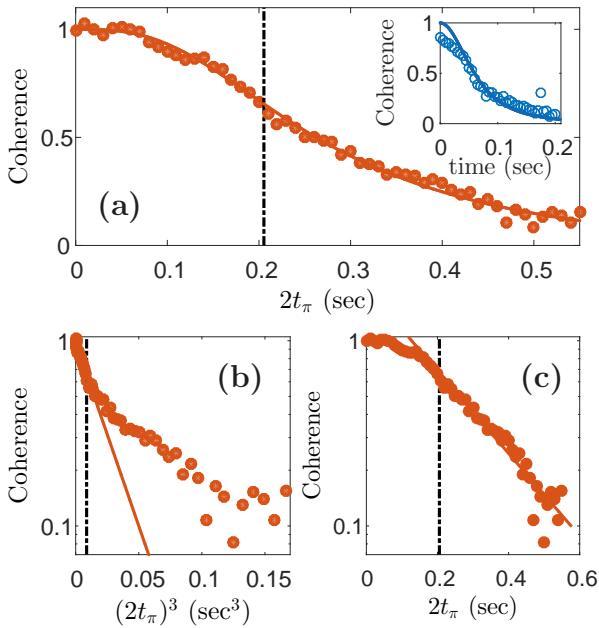}
\put(40,165){\large \textbf{(a)}}
\put(200,100){\large \textbf{(c)}}
\put(85,100){\large \textbf{(b)}}
\end{overpic}
\caption{Echo measurement, with $\Gamma_\text{coll} t_\text{tr} \approx 1$, showing a transition of $\alpha$. (a) The data is fit to the function given in Eq.~\eqref{eq:Transition_func}, yielding $2t_\text{tr} \approx 205$~ms [dashed vertical lines in (a-c)], using $\Gamma_\text{coll}=5.1~(3)~\text{s}^{-1}$ obtained from (c) as a set parameter. The inset shows a Ramsey experiment performed under the same experimental conditions, fitted (solid line) to Eq.~\eqref{eq:Meschede}, yielding a Ramsey time $\tau=76~(3)$~ms. (b) Same data with time axis rescaled to $(2t_\pi)^3$, showing the transition clearly on a semi-log scale. The linear behavior at short times (solid line represents linear fit) indicates the $e^{-(2t_\pi)^3}$ dependence. (c) The same data on a semi-log scale as a function of $2t_\pi$. The linear dependence at long times (solid line represents linear fit) indicates the $e^{-2t_\pi}$ decay.}
\label{fig:fig4}
\end{figure}

We measure the echo coherence decay with a moderate collision rate ($\Gamma_\text{coll} t_\text{tr} \approx 1$). The resultant coherence is summarized in Fig.~\ref{fig:fig4}. The transition of $\alpha$ is evident from the fit to Eq.~\eqref{eq:Transition_func}. An indication for the $\alpha = 3$ decay is shown in Fig.~\ref{fig:fig4}(b), where we plot the echo revival amplitudes in a logarithmic scale against $(2t_\pi)^3$. Fig.~\ref{fig:fig4}(c) shows the same data in a semi-logarithmic scale as a function of $2t_\pi$. Here the linear dependence at long times indicates the $\alpha = 1$ decay. A similar transition was observed for warm molecular gases in the limit of soft collisions~\cite{Schmidt1973}.

More quantitatively, extracting $t_\text{tr}$ from the echo decay using Eq.~\eqref{eq:Transition_func}, and $\tau$ from the Ramsey decay under the same experimental conditions, we can evaluate the softness by the definition of the transition time $s = (t_\text{tr}/\tau)^2$. Fitting the echo decay of Eq.~\eqref{eq:Transition_func} to the data of Fig.~\ref{fig:fig4}, using $\Gamma_\text{coll}=5.1~(3)~\text{s}^{-1}$ obtained from fitting the long-time decay of Fig.~\ref{fig:fig4}(c), we get $2t_\text{tr}=205~(17)~$ms~\footnote{To evaluate the error $2\Delta t_\text{tr}$ we use the value $\Gamma_\text{coll}\pm\Delta\Gamma_\text{coll}$ as a fitting parameter. The upper $1\sigma$ confidence bound on $t_\text{tr}$ is that which is obtained for $\Gamma_\text{coll}+\Delta\Gamma_\text{coll}$ and the lower bound is that which is obtained for $\Gamma_\text{coll}-\Delta\Gamma_\text{coll}$}. This, in addition with a Ramsey experiment under the same experimental conditions [inset of Fig.~\ref{fig:fig4}(a)] that gives $\tau=76~(3)$~ms, yields $s = (t_\text{tr}/\tau)^2=1.8~(3)$. The agreement with the theoretical value of 2.5 is not as good as for the measurement described in the previous section.

\begin{figure}
\centering
\begin{overpic}[width=\linewidth]{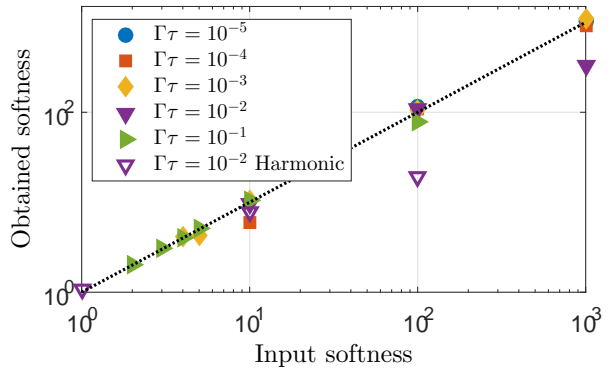}
\end{overpic}
\caption{Numerical investigation of the effectiveness of using Eq.~\eqref{eq:Transition_func} as a fitting function for extracting the softness for a Gaussian energy distribution (full symbols) and the energy distribution corresponding to a 3D harmonic potential (empty symbols). The simulation calculates, for a predetermined input softness, a Ramsey decay curve as well as a full echo coherence decay curve. To approximate the obtained softness, we simultaneously fit the Ramsey decay to Eq.~\eqref{eq:Gumbel}, and the echo decay to Eq.~\eqref{eq:Transition_func}, with three fitting parameters: $\Gamma$, $\tau$ and $s$. Black dotted line is where the obtained softness is equal to the real one.}
\label{fig:fig5}
\end{figure}

We investigate the use of the interpolation function of Eq.~\eqref{eq:Transition_func} as a fitting function for extracting the softness using the Monte Carlo simulations described previously. We find that for a Gaussian energy distribution the method is quite precise. Fig~\ref{fig:fig5} presents a summary of the softness obtained for a Gaussian energy distribution (full symbols) by fitting Eq.~\eqref{eq:Transition_func} to the numerically simulated decay of coherence as a function of the input softness for different values of $\Gamma\tau$. For most cases, the output softness is very close to the input softness. The relative error is $<10\%$ as long as $\Gamma_\text{coll}t_\text{tr}$ is within the range $0.1<\Gamma_\text{coll}t_\text{tr}<10$. Outside this range, the value of the coherence at the transition time $C_\text{E}(2t_\text{tr})\approx \exp\left(-2\Gamma_\text{coll}t_\text{tr}\right)$ is either too high or too low for the fitting procedure to accurately extract the transition time. For the case of an energy distribution of a 3D harmonic trap similar to the one we have in the experiment the situation is different [Fig.~\ref{fig:fig5}, empty symbols]. The approximate solutions for the Ramsey [Eq.~\eqref{eq:Gumbel}] and echo [Eq.~\eqref{eq:Transition_func}] fit only qualitatively, yielding significant errors similar to the ones observed in the experiment.

\subsection*{Summary and outlook}

We have measured the softness of ultracold elastic atomic collisions using a combination of two spectroscopic methods: a measurement of the energy decorrelation rate obtained from the collisional narrowing of a Ramsey experiment with a high collision rate, and a direct measurement of the collision rate obtained from an echo experiment with a low collision rate. The obtained collisional softness is 2.5 (3) - in excellent agreement with the value obtained by molecular dynamics simulations.  We have also demonstrated a transition in the functional decay of the echo coherence, from $\exp\left(-t^3\right)$ at short times, to $\exp\left(-t\right)$, at long times~\cite{Schmidt1973}. This transition occurs at a softness-dependent time $t_\text{tr}=\tau/\sqrt{s}$. We show that this transition in the functional decay can be used to approximate the softness at a single, intermediate density regime. 

Our results validate the spectroscopic method, allowing for its use in measuring the softness of other, non-trivial, collisional processes such as extensions to higher temperature involving the inclusion of more partial waves into the scattering process, fermionic collisions and thermalization~\cite{Brantut2013}, inter-species hybrid collisions~\cite{Barbut2014,Delehaye2015} and atom-ion collisions~\cite{Meir2016}. It can also be useful as a tool in simulating the efficiency of evaporative cooling~\cite{Foot1996}, the investigation of high-density atom interferometers~\cite{Fogarty2013} and slow and stored light~\cite{firstenberg2013colloquium}. Our methods may allow measurements of a generalized softness parameter for other two-level quantum systems with discrete spectral jumps~\cite{koch2007model, wodkiewicz1984noise,ambrose1991fluorescence,uren19851}.

Our work can be extended to warm vapor systems where the bare Ramsey time is dominated by Doppler broadening and the presence of a buffer gas induces collisional narrowing~\cite{Dicke1953}. As in our system, the effect of collisional softness on the Ramsey signal is in itself very small and challenging to detect~\cite{Rothberg081984,firstenberg2013colloquium}. Combining Raman Rabi, Ramsey and echo spectroscopy at high and low collisional-rate regimes can provide an accurate measure of the collisional softness for different buffer gases. In this case, it is possible to change $\Gamma_\text{coll}\tau$ by simply altering the angle between the Raman beams without actively changing the density~\cite{firstenberg2013colloquium}.

\begin{acknowledgments}
The authors would like to thank Yoav Sagi, Hagai Edri and Noam Matzliah for discussions. This work was supported in part by the ICore Israeli excellence center Circle of light and the Weizmann Institute Texas A\&M collaboration program.
\end{acknowledgments}


\bibliographystyle{apsrev4-1}
\bibliography{Collisions_bib}
\end{document}